\documentclass[aps,pra,reprint,amssymb,superscriptaddress]{revtex4-2} 
\usepackage{amsmath,amssymb,amsfonts}
\usepackage{graphicx}
\usepackage{dcolumn}
\usepackage{bm}
\usepackage{booktabs}
\usepackage[colorlinks=true,citecolor=blue,linkcolor=blue,urlcolor=blue]{hyperref}
\usepackage[mathlines]{lineno}

\usepackage{float}
\makeatletter
\let\newfloat\newfloat@ltx
\makeatother
\usepackage{algorithm}
\usepackage{algpseudocode}
\usepackage{dsfont}
\usepackage{orcidlink}
\usepackage[utf8]{inputenc}
\usepackage[T1]{fontenc}

\usepackage{braket}
\usepackage[dvipsnames]{xcolor}
\usepackage{tabularray}

\definecolor{tms}{rgb}{1,0,0}

\begin{document}

\title{Exploring the Relaxation Landscape of a 2D Quantum Magnet on a 256-Qubit Processor}


 \author{Tiago Mendes-Santos\textsuperscript{*}\, \orcidlink{0000-0001-6827-5260}}
 \email{tiago.mendes-santos@pasqal.com}
 \author{Joseph Vovrosh\textsuperscript{*}\,\orcidlink{0000-0002-1799-2830}}
 \email{joseph.vovrosh@pasqal.com}
 \author{Sergi Juli\`{a}-Farr\'e\textsuperscript{*}\,\orcidlink{0000-0003-4034-5786}}
  \email{sergi.julia-farre@pasqal.com}
 \author{Dorian Claveau\textsuperscript{*}}
 \author{Guillaume Villaret~\orcidlink{0000-0002-3898-8646}}
 \author{Lucas Béguin}
  \author{Lucas Leclerc}
   \let\comma,
 \affiliation{Pasqal, 24 rue Emile Baudot - 91120 Palaiseau,  Paris, France}

 \author{Laurin Brunner}
 \affiliation{Theoretical Physics III, Center for Electronic Correlations and Magnetism, Institute of Physics, University of Augsburg, D-86135 Augsburg, Germany}
 \author{Wladislaw Krinitsin\,\orcidlink{0009-0009-2169-1246}}
 \let\comma,
 \affiliation{Institute of Quantum Control (PGI-8), Forschungszentrum Jülich, D-52425 Jülich, Germany}
 \affiliation{Faculty of Informatics and Data Science, University of Regensburg, D-93053 Regensburg, Germany}

\author{Matthias Hecker}
\author{Fergus Hayes\,\orcidlink{0000-0001-7628-3826}}
\author{Boris Albrecht~\orcidlink{0000-0003-0733-2676}}
\author{Lilian Bourachot~\orcidlink{0009-0002-5860-9903}}
\author{Cl\'{e}mence Briosne-Frejaville~\orcidlink{0009-0007-2615-5050}}
\author{Antoine Cornillot\,\orcidlink{0009-0004-7228-541X}}
\author{Julius de Hond\,\orcidlink{0000-0003-2217-934X}}
\author{Djibril Diallo~\orcidlink{0009-0007-8393-9329}}
\author{Cl\'{e}ment Dupays~\orcidlink{0009-0007-1944-7223}}
\author{Robin Dupont~\orcidlink{0009-0003-4541-5146}}
\author{Thomas Eritzpokhoff~\orcidlink{0009-0000-3899-8564}}
\author{Lo\"ic Henriet~\orcidlink{0000-0003-3108-0595}}
\author{Lucas Lassabli\`ere~\orcidlink{0000-0001-8081-1054}}
\author{Arvid Lindberg~\orcidlink{0000-0001-8714-8662}}
\author{Yohann Machu~\orcidlink{0009-0007-6766-6439}}
\author{Hadriel Mamann~\orcidlink{0009-0002-3832-6471}}
\author{Thomas Pansiot~\orcidlink{0009-0001-2099-0043}}
\author{Julien Ripoll~\orcidlink{0009-0004-5282-9942}}
\author{Bruno Ximenez~\orcidlink{0009-0006-8985-1355}}
\author{Henrique Silv\'{e}rio~\orcidlink{0000-0003-0416-5518}}
 \affiliation{Pasqal, 24 rue Emile Baudot - 91120 Palaiseau,  Paris, France}

 \author{Joseph Tindall\,\orcidlink{0000-0003-1335-8637}}
 \let\comma,
 \affiliation{Center for Computational Quantum Physics, Flatiron Institute, 162 Fifth Avenue, New York,
 New York 10010, USA}
 \author{Markus Schmitt\,\orcidlink{0000-0003-2223-8696}}
 \let\comma,
 \affiliation{Institute of Quantum Control (PGI-8), Forschungszentrum Jülich, D-52425 Jülich, Germany}
 \affiliation{Faculty of Informatics and Data Science, University of Regensburg, D-93053 Regensburg, Germany}
 \author{Markus Heyl\,\orcidlink{0000-0002-7126-1836}}
 \affiliation{Theoretical Physics III, Center for Electronic Correlations and Magnetism, Institute of Physics, University of Augsburg, D-86135 Augsburg, Germany}
 \affiliation{Centre for Advanced Analytics and Predictive Sciences (CAAPS), University of Augsburg, Universitätsstr. 12a, 86159 Augsburg, Germany}

 \author{Adrien Signoles~\orcidlink{0000-0001-7822-9444}}
 \author{Constantin Dalyac\,\orcidlink{0000-0002-0339-6421}}
\let\comma,
 \affiliation{Pasqal, 24 rue Emile Baudot - 91120 Palaiseau,  Paris, France}
  \author{Antoine Browaeys\,\orcidlink{0000-0001-9941-8869}}
  \let\comma,
 \affiliation{Université Paris-Saclay, Institut d’Optique Graduate School,
 CNRS, Laboratoire Charles Fabry, 91127 Palaiseau Cedex, France}
 \author{Alexandre Dauphin\,\orcidlink{0000-0003-4996-2561}}
 \email{alexandre.dauphin@pasqal.com}
  \let\comma,
 \affiliation{Pasqal, 24 rue Emile Baudot - 91120 Palaiseau,  Paris, France}

\begin{abstract}
How quantum matter relaxes far from equilibrium is a central open problem in many-body physics, and one for which analog quantum simulators are well positioned to move from confirming theory to discovering new physics. Here, we use a two-dimensional Rydberg atom array of 256 qubits to map the relaxation landscape of the two-dimensional transverse-field Ising model across its phase diagram. Beyond the expected rapid thermalization, we identify two further regimes.
The first is a prethermal regime whose dynamics are governed by an effective XY model.
The second, and most unexpected, is a crossover regime characterized by a slowdown in relaxation. This slowdown occurs precisely where state-of-the-art classical tensor-network methods lose control at late times, whereas the quantum simulation remains consistent across system sizes. These results establish Rydberg atom arrays as a platform for scientific discovery in nonequilibrium quantum many-body dynamics.
\end{abstract}

\maketitle

Analog quantum simulators have become powerful platforms for probing nonequilibrium many-body physics~\cite{browaeys_many-body_2020,henriet_quantum_2020}, and have revealed universal dynamical phenomena such as dynamical phase transitions~\cite{jurcevic_direct_2017,flaschner_observation_2018,zhang_observation_2017}, scaling across quantum critical points~\cite{scholl_quantum_2021,ebadi_quantum_2021}, and emergent hydrodynamic behavior~\cite{joshi_observing_2022,zu_emergent_2021,wienand_emergence_2024}. So far, most of these results have reproduced effects already anticipated by theory. Their greater promise is to act as genuine discovery platforms, revealing physics that theory did not foresee, a frontier that only a few pioneering studies have begun to reach~\cite{bernien_probing_2017,de_leseleuc_observation_2019,manovitz_quantum_2025,darbha_probing_2025}.

Far from equilibrium, relaxation pathways are not unique: interacting quantum systems may fail to thermalize quickly and instead settle into long-lived transient states beyond the reach of conventional equilibrium statistical mechanics~\cite{polkovnikov_nonequilibrium_2011,eisert_quantum_2015,ho_quantum_2023}. Prominent examples are prethermalization, in which the system is trapped in a nonthermal state that can persist for exponentially long times, with rich many-body structure~\cite{abanin_rigorous_2017,mori_thermalization_2018,ho_quantum_2023}, and quantum many-body scars, which produce atypical dynamical revivals from special initial states~\cite{turner_weak_2018}. How such transient regimes form, and how they eventually give rise to thermalization, has become a central question especially in two-dimensional quantum matter, where it remains far less understood than in one dimension. 

\begin{figure*}
    \centering
    \includegraphics[width=\linewidth]{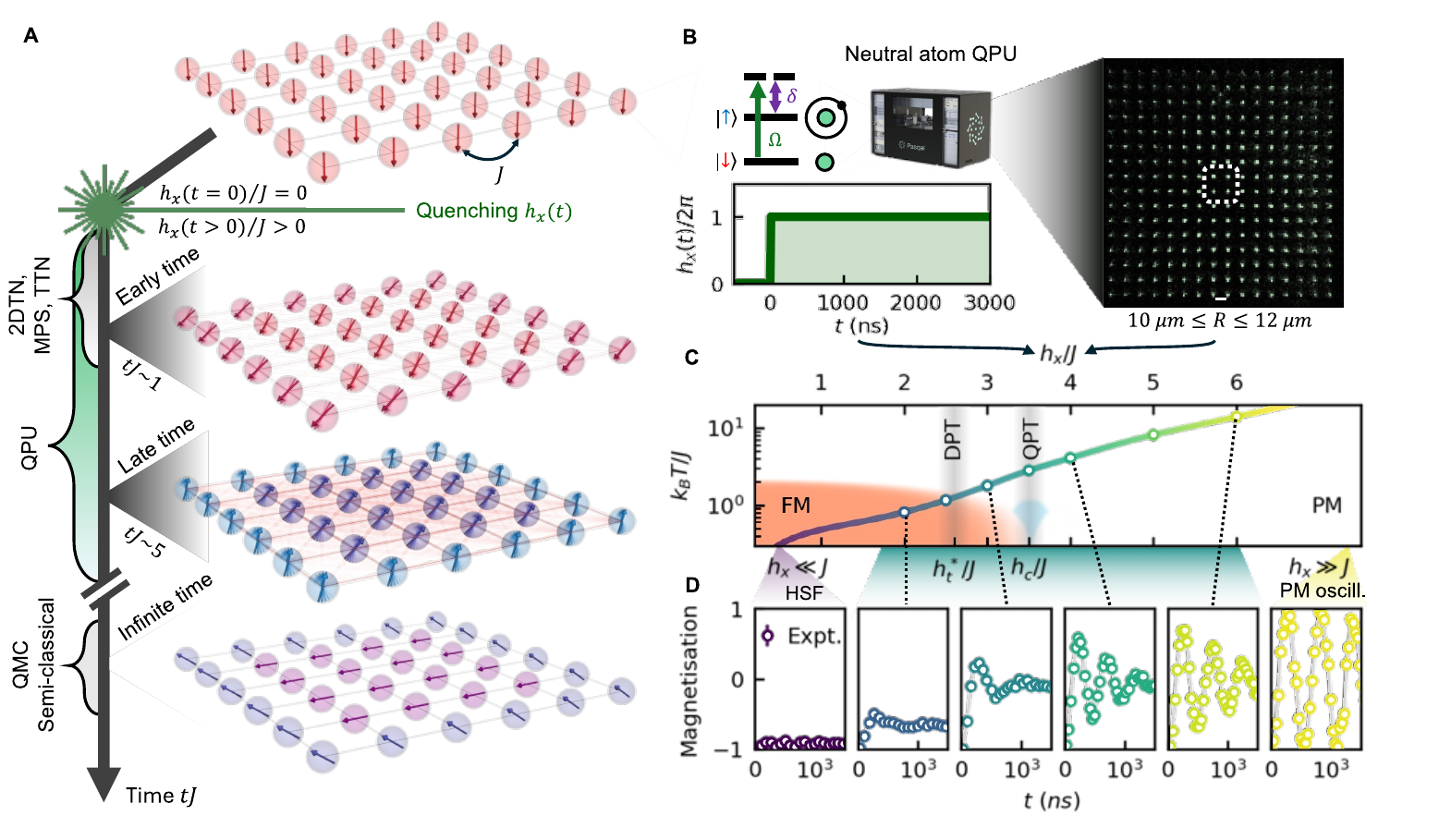}
    \caption{\textbf{Post-quench dynamics in the 2D TFIM.} 
(A) We consider quantum simulations of post-quench dynamics described by the 2D TFIM. The qubits are initialized in a fully polarized state, and we study the dynamics following quenches to different values of $h_x$, while $h_z^i$ is chosen such that the longitudinal field in the bulk satisfies $h_z = 0$ (see text). 
(B) In particular, we consider quantum simulators based on Rydberg atoms arranged in square arrays containing $L \times L$ qubits, with $J$ denoting the nearest-neighbor Rydberg-Rydberg interaction. Control over the pulse shapes of the Rabi frequency, $\Omega(t)$, and the detuning, $\delta(t)$, as well as the relative distance between atoms, allows one to program quenches to various values of $h_x/J$. 
(C) We focus on the non-perturbative regime of post-quench dynamics described by the ferromagnetic 2D TFIM phase diagram, which features ferromagnetic (FM) and paramagnetic (PM) phases, and a transition line between those phases. We highlight in the phase diagram the effective temperature associated with the steady state, which is obtained by comparing the energy of the initial state with that of a thermal state computed using QMC simulations (see Methods). The crossing of this line with the phase-transition line allows us to obtain a thermal prediction for the DPT, $h_t^*$. In particular, we consider the values  $h_x/J \in [2.0, 6.0]$. Examples of the magnetization dynamics are shown in panel (D).}
    \label{fig:Fig1}
\end{figure*}

Here we address one such case: the post-quench dynamics of the two-dimensional transverse-field Ising model (TFIM) on a programmable Rydberg-based analog quantum processing unit (QPU) of 256 qubits. Starting from the fully polarized product state $\ket{\downarrow\cdots\downarrow}$, we quench to constant transverse field $h_x$ and interaction $J$, and tune the ratio $h_x/J$ to scan across the model's phase diagram (Fig.~\ref{fig:Fig1}). The two-dimensional ferromagnetic TFIM is a demanding testbed: its equilibrium phase diagram hosts both a thermal and a quantum phase transition, and quenches from the polarized state are expected to exhibit a dynamical phase transition (DPT)~\cite{mondaini_eigenstate_2016}. 
How the system relaxes in the vicinity of the DPT remains largely open, and is precisely the regime where two-dimensional dynamics become hardest to capture by classical means.

We find that this initial state gives rise to a surprisingly rich relaxation landscape. When the transverse field is comparable to the interaction strength, $h_x \sim 2J$, the system thermalizes rapidly, as expected. At large $h_x/J$, where the transverse field dominates, it instead enters a long-lived prethermal regime described by an effective Floquet Hamiltonian~\cite{abanin_rigorous_2017,ho_quantum_2023}. Interestingly, unlike the Rydberg-blockade parameter regime, which leads to an effective PXP-type description~\cite{bernien_probing_2017,darbha_probing_2025}, the corresponding effective description here is given by a $U(1)$ XY-type model and is characterized by a strong dynamical buildup of connected correlations.
Additionally, between the thermal and prethermal quench limits, we discover an extended crossover regime in which thermalization slows down. Benchmarking against tensor-network methods confirms these regimes at short times. At late times, however, classical methods lose control precisely in this slow-relaxation regime, whereas our quantum simulations remain consistent across system sizes.

\section*{Quantum simulations of post-quench dynamics in the 2D TFIM }
Our quantum simulations are performed with a two-dimensional square array of $N=L\times L$ trapped $^{87}$Rb atoms where local qubits are encoded in the atomic ground and Rydberg states, denoted by $\ket{\downarrow}$ and $\ket{\uparrow}$, as illustrated in Fig.~\ref{fig:Fig1}\textbf{B}. To simulate a many-body spin Hamiltonian, $\hat{H}_\textrm{QPU}$, the qubit states are coupled through a two-photon excitation with global Rabi frequency $\Omega$ and detuning $\delta$, and each qubit pair at a distance $r_{ij}$ interacts via short-range van der Waals interactions $U_{ij}=U(R/r_{ij})^6$; here, $U$ and $R$ are the nearest-neighbor interaction and distance, respectively. This setup allows us to simulate the post-quench dynamics of a fully polarized initial state $\ket{\downarrow\cdots\downarrow}$ (see Fig.~\ref{fig:Fig1}A) under a constant short-range 2D TFIM Hamiltonian
\begin{equation}\label{eq:ham_qpu_}
\begin{split}
    \hat H_{\text{TFIM}} & = -\sum_{i<j} J_{ij} \hat \sigma^z_i\hat \sigma^z_j -
    h_x \sum_i \hat\sigma^x_i,
 \end{split}
\end{equation}
where $h_x=\Omega/2$ is the transverse field, and $J_{ij}=J(R/r_{ij})^6$ denotes the short-range ferromagnetic interaction, with $J=U/4\geq 0$. In the Methods, we provide details on the mapping. In particular, the zero longitudinal field condition is achieved  in the bulk of the atomic array with a global $\delta$, leaving residual site-dependent edge terms $\propto \hat{\sigma}^z_i$ in $\hat{H}_\textrm{QPU}$.
Experimentally, working with angular frequencies ($\hbar=1$) the timescale is set by $h_x/(2\pi)=1\,\textrm{MHz}$. Varying $R$ tunes the ratio $h_x/J$, and thus the dynamical regime explored during the evolution time $t$. We express the latter in ns or $1/J$ units, related by the ratio $h_x/J$.
The measurement of observables is then performed using $300$ experimental shots per time step, retaining only shots with $<1.5\%$ defects leaving $\sim 250$. SPAM mitigation is applied (see Methods), and observables are averaged over equivalent qubits/qubit pairs. Error bars are obtained from bootstrap resampling.

Here, we focus on the regime $h_x > J$.
In this case, the quench dynamics exhibits a DPT in the asymptotic late-time order parameter. For quenches to transverse fields below a certain threshold, $h_x < h_t^*$, the magnetization remains finite at long times, showing that the system retains the $\mathbb{Z}_2$ symmetry breaking of the initial fully polarized phase. For stronger quenches, $h_x > h_t^*$, this memory is lost: the magnetization either relaxes to zero or oscillates around zero with a vanishing time average, consistent with restoration of the Ising $\mathbb{Z}_2$ symmetry.
This behavior has a natural interpretation in terms of the eigenstate thermalization hypothesis (ETH). Under ETH, the post-quench steady state is characterized by a thermal ensemble whose effective temperature is fixed by the initial state. A thermal estimation of $h_t^*$ is set by the point at which this effective temperature crosses the finite-temperature critical line of the 2D TFIM phase diagram, as illustrated in Fig.~\ref{fig:Fig1}\textbf{C}.
It is worth noting that, although here we consider DPTs in the late-time order parameter, singular dynamical behavior can also appear in the transient real-time evolution of Loschmidt echoes, which is also commonly referred to as DPTs \cite{halimeh_prethermalization_2017,zunkovic_dynamical_2018,hashizume_dynamical_2022}.

As a first QPU result, the magnetization $\langle \hat{\sigma}^{z}_{i}(t)\rangle$ averaged over the central sites, is consistent with such a DPT at $L=16$ (Fig.~\ref{fig:Fig1}D). In the PM phase at large $h_{x}/J$ we
observe the dynamical restoration of the $\mathbb{Z}_{2}$ symmetry through oscillations of the bulk magnetization; these are increasingly damped as $h_{x}/J$ decreases, and in the limit $h_{x}\gg J$ reduce to collapses and revivals close to single-atom Rabi oscillations. In the FM region at small $h_{x}/J$ the magnetization instead retains the sign of the initial state. The crossover between the two regimes occurs between $h_{x}/J = 2.5$ and $3.0$, consistent with the thermal estimate $h_{t}^{*}/J\approx 2.6$ for the QPU Hamiltonian, i.e.,  $\hat{H}_\textrm{QPU}$ (Methods).

\begin{figure}
    \centering
    \includegraphics[width=\linewidth]{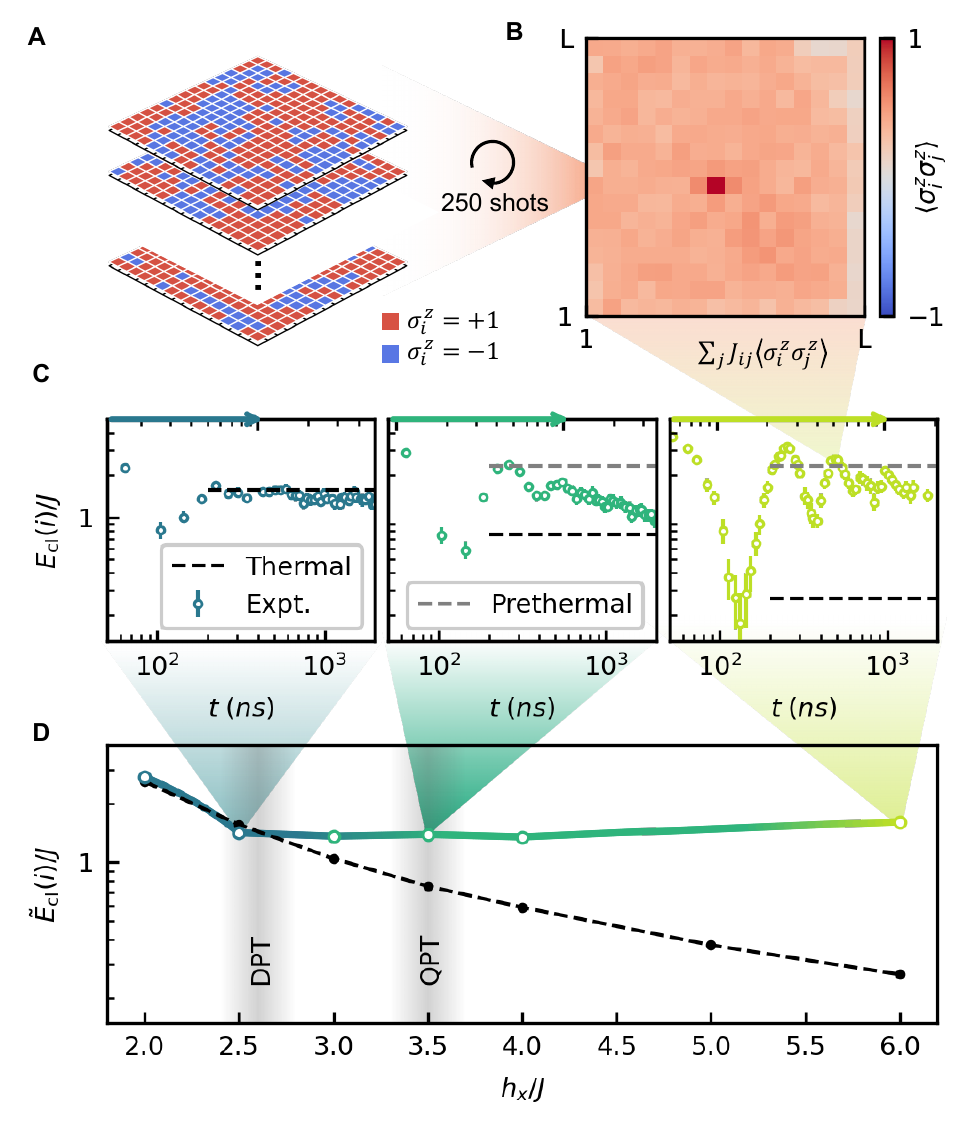}
    \caption{\textbf{Distinct regimes of nonequilibrium relaxation in the 2D TFIM ($16\times 16$ array).} 
(A) Site-resolved projective measurements of $\hat\sigma_i^z$ enable the construction of (B) the correlation matrix, $\langle \hat\sigma_i^z \hat\sigma_j^z \rangle$. 
(C) Time evolution of the local classical energy $E_{\mathrm{cl}}(i,t)$ averaged over central sites for quenches to (left) $h_x/J = 2.5$, (middle) $h_x/J = 3.5$, and (right) $h_x/J = 6.0$. The dashed black line indicates the thermal expectation value, and the gray dashed line indicates the prethermal expectation value. The arrows in the upper part of each panel indicate the time interval of $t = 1/J$.
(D) Time-averaged classical energy $\tilde{E}_{cl}(i)$ as a function of $h_x/J$, together with the corresponding thermal expectation values. Error bars, which are obtained with bootstrapping, are smaller than the point sizes.}
    \label{fig:Fig2}
\end{figure}

\section*{Relaxation landscape}

While the connection to the TFIM phase diagram predicts infinite-time observables, the ability of the QPU to monitor the intermediate-time dynamics is essential to reveal purely nonequilibrium phenomena. Having established qualitatively distinct regimes in the magnetization dynamics, we now go beyond this local-observable diagnostic and examine the underlying relaxation mechanism by analyzing two-body correlations in the system. To this aim, we use the same projective measurements on the $L=16$ system to extract the time-dependent classical (Ising) energy of central site $i$ $E_{\mathrm{cl}}(i,t) = \sum_j J_{ij} \langle \hat{\sigma}_i^z \hat{\sigma}_j^z \rangle(t)$,
i.e. the diagonal part of the local energy, built only from $\hat{\sigma}^{z}$ correlations (see Figs.~\ref{fig:Fig2}{\bf A} and {\bf B}). 
Its evolution for different $h_{x}/J$ reveals qualitatively distinct relaxation behaviors of the correlations, illustrated in Figs.~\ref{fig:Fig2}{\bf C} for three representative values of $h_x/J$; throughout, we measure time in units of the interaction scale $1/J$.

For (i) quenches deep in the PM phase, e.g., $h_x/J = 6.0$, the correlation dynamics contrast sharply with the magnetization: instead of relaxing or oscillating around their thermal value (black dashed line in  Figs.~\ref{fig:Fig2}{\bf C}), the correlations are sustained at a higher value, suggesting the emergence of a prethermal mechanism. For (ii) quenches in the PM phase but for lower values of $h_x/J$, e.g., $h_x/J = 3.5$, relaxation toward thermal equilibrium occurs but is markedly slow, with the correlations remaining 
far from their thermal value  beyond the characteristic relaxation time $t\approx 1/J$. Finally, for (iii) a quench in the vicinity of the DPT, $h_x/J=2.5$, the correlations thermalize rapidly, i.e., at time scales $t \approx 1/J$. We now discuss each of these regimes in turn.

\subsection*{Floquet prethermalization at large transverse field}
We attribute the absence of thermalization in the correlations of Fig.~\ref{fig:Fig2}{\bf C} at $h_x/J=6.0$ to the emergence of an effective Floquet Hamiltonian at large transverse fields $h_x \gg J$~\cite{abanin_rigorous_2017,ho_quantum_2023}.
In the rotating frame defined by the transverse field, the Ising coupling becomes time-periodic in $T_\textrm{F}=\pi/h_x$, so the system behaves as an effectively (self-)driven Floquet system and, on a timescale $\mathcal{O}(1/J)$, first relaxes to a prethermal plateau described by the thermal properties of the leading-order effective Hamiltonian. Time-averaging over one period yields a $U(1)$-symmetric rotated XY model, i.e. a ZY model (Methods)
\begin{equation}
\hat{H}_{ZY} = \frac{1}{2} \sum_{i<j} J_{ij} \left( \hat{\sigma}_i^z \hat{\sigma}_j^z + \hat{\sigma}_i^y \hat{\sigma}_j^y \right).
\label{eq:ZYmodel}
\end{equation}
Exploiting the $U(1)$ symmetry of $\hat{H}_{\mathrm{ZY}}$ (Methods), we analytically extract a prethermal value of $E_{\mathrm{cl}}(i,t)$  (gray dashed line in Figs.~\ref{fig:Fig2}{\bf C}) in qualitative agreement with the QPU data.
Corrections from higher-order terms in the expansion are expected, and the slightly smaller correlations measured on the QPU are also compatible with the finite coherence time of the device (see Methods for a classical benchmark at $L=5$). At later times this plateau is expected to melt on a timescale growing exponentially in $h_{x}/J$, ultimately leading to thermalization.

\subsection*{Slowdown of thermalization in the crossover regime}
Although prethermalization governed by $\hat{H}_{\textrm{ZY}}$ provides an accurate description of the dynamics deep in the PM region, this picture progressively breaks down as one moves away from this perturbative regime. This behavior is illustrated in Fig.~\ref{fig:Fig2}{\bf C}: for $h_x/J = 3.5$, the correlations do not form a clear plateau; instead, we observe an unexpected regime in which they decay slowly toward their thermal value.
While we do not yet have a clear explanation for this discovery, we conjecture that the slow thermalization observed in this regime results from the interplay of two effects.
First, the slowdown may signal the breakdown of the effective  leading-order Hamiltonian description  in terms of $\hat{H}_{\textrm{ZY}}$, which constrains the dynamics only in the perturbative limit $h_x \gg J$. Away from this regime, additional terms in the effective Hamiltonian become relevant.
Such higher-order corrections may play an important role in restoring thermalization.  
Second, for quenches close to the thermal transition, the thermal prediction controlling the long-time dynamics is also consistent with critical slowing down~\cite{mitra_macroproperties_2025}. 
In the method section, we perform a small system size analysis with state vector simulations on a $5 \times 5$ lattice with and without noise. We recover the same qualitative behavior and observe that both the system size and the noise can affect the slope of the slowdown. A systematic characterization of these contributions and the effect of the platform's noise is the next natural step that we plan to explore in the future. It would require either larger scale accurate numerics or access to the next generation of QPUs with lower noises.

\subsection*{Thermalization in the ferromagnetic phase}

In the FM region of the phase diagram, i.e., for $h_x/J = 2.0$ and $2.5$, the results for $E_{\mathrm{cl}}(i,t)$ at times $t > 1/J$ are well described by the corresponding thermal values of $\hat{H}_\textrm{QPU}$, defined by the effective temperature fixed by the initial fully polarized state (see Figs.~\ref{fig:Fig2}\textbf{C}).
 One important remark is that for those quenches the long-time behavior of the bulk magnetization stabilizes to a finite value, which is consistent with a thermalization into a symmetry-broken sector \cite{fratus_eigenstate_2015}. This is in contrast with  quenches to $h_x > h_t^*$, where the symmetry  breaking of the initial state is dynamically restored, as manifested by oscillations of the bulk magnetization around zero (see Fig. \ref{fig:Fig1} {\bf D}).

Finally, to further characterize the qualitative changes between the different relaxation regimes, we also consider time-averaged correlations $\tilde{E}_{\mathrm{cl}}(i)=(1/\delta t)\int_{t_{\min}}^{t_{\max}} E_{\mathrm{cl}}(i,t)dt$, with $t_{\min}J\approx1$ and $t_{\max}J\approx2$ (with $\delta t=t_{\max}-t_{\min}$), as a function of $h_x/J$. 
As shown in Fig.~\ref{fig:Fig2}{\bf D}, the monotonic  thermal prediction contrasts with a dip in the  QPU data: around the crossover, the time-averaged correlations fall below both the high-field prethermal plateau and the low-field thermal value.
This picture shows the qualitative change of behavior between the rapid  thermalization and prethermalization regime. However, establishing a quantitative characterization of  $\tilde{E}_{\mathrm{cl}}$ in the crossover regime would require further system-size and finite-time scaling analyses of the QPU data shown in Fig.~\ref{fig:Fig2}\textbf{D}, which go beyond the current implementation (see Methods for a numerical analysis in $L=5$).

\begin{figure}
    \includegraphics[width=1\columnwidth]{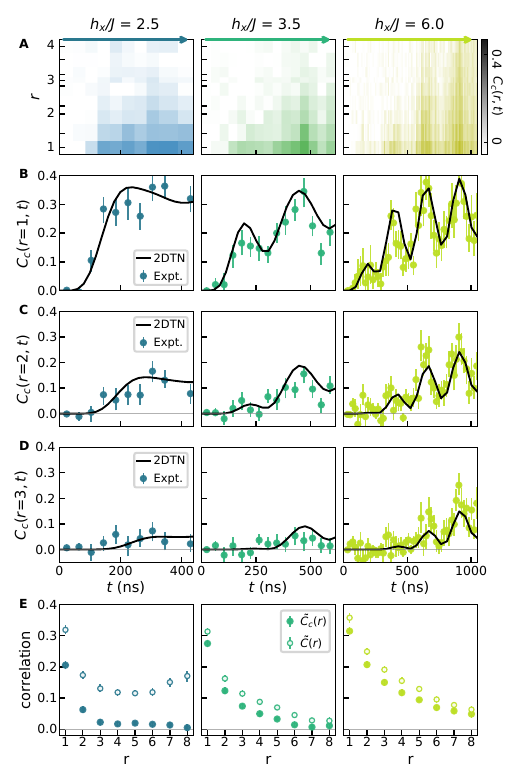}
    \caption{\textbf{Spatiotemporal buildup of connected correlations in the 2D TFIM ($16\times 16$ array)} for $h_x/J=2.5, 3.5, 6$ (left to right).  (A) Space-time map of the connected correlator $C_c(r,t)$ (main text and Methods). Arrows on top axis show the early-time window until $tJ=1$. The right-hand grayscale colorbar provides a common intensity normalization shared by all three heatmaps. (B-D) Time trace of $C_c(r,t)$ for first (B), second (C) and third (D) neighbors along a lattice axis. Classical simulations based on 2DTN are shown for comparison (solid lines). Panel (E) shows the time-averaged, $tJ\in[1,2]$, correlations and connected correlations as function of the distance $r$.}
    \label{fig:Fig3}
\end{figure}

\section*{Signatures of the relaxation regimes in the correlation spread}

Building on the results of the different relaxation regimes
of the 2D TFIM, we now examine 
the spatiotemporal buildup of equal-time connected correlations, $C_c(i,j,t)= \langle \hat{\sigma}_i^z \hat{\sigma}_j^z \rangle(t) - \langle \hat{\sigma}_i^z \rangle \langle \hat{\sigma}_j^z \rangle(t)$. To mitigate edge effects, we consider the average $C_c(r,t)$ over Euclidean shells at distance $r$ around the four central sites in the $L=16$ lattice.

\subsection*{Early-time  correlation spread}
Figure~\ref{fig:Fig3}\textbf{A} shows the propagation $C_c(r,t)$ over time, which reveals qualitatively distinct behavior across the relaxation regimes described in Fig.~\ref{fig:Fig2}. To examine this in more detail, in Figs.~\ref{fig:Fig3}\textbf{B-D} we plot the first three shells along a line. As a certification, we show quantitative agreement of these QPU results with fully converged two-dimensional tensor network (2DTN) simulations, which provide numerically exact reference data at early times up to $tJ \approx 1$, despite the large system size $L=16$ (Methods).

For $h_x/J = 6.0$, we observe pronounced temporal oscillations in all $C_{c}(r,t)$, with an expected period $T_\textrm{F}/2$ (250 ns) that is halved with respect to the one of the paramagnetic oscillations of the magnetization in Fig.~\ref{fig:Fig1}D. This behavior indicates that the build-up of correlations, driven by the interaction scale $J$, is strongly constrained by the faster scale induced by the transverse field $h_x$. This regime is thus consistent with the effective ZY model and emergent $U(1)$ symmetry at $h_x\gg J$. 

At the intermediate $h_x/J=3.5$, we also observe similar oscillations with approximately $T_\textrm{F}/2$ period. However, a clear separation of energy scales is lost: only two oscillations occur within $tJ\lesssim 1$, and already the first oscillation is accompanied by a sizable nearest-neighbor correlation. These competing energy scales are compatible with the later relaxation dynamics, in which despite the transverse field brings the system into the PM regime, it is not strong enough to stabilize a prethermal state.

Finally, for $h_x/J = 2.5$, we observe a rapid and essentially featureless spread of correlations. We also note that longer-range correlations, $C_{c}(r>2,t)$, remain strongly suppressed over time, consistent with fast thermalization within a symmetry-broken sector; that is, the system relaxes toward a state with $\langle \hat\sigma_i^z \rangle \neq 0$, which in turn reduces the magnitude of $C_{c}(r,t)$.

\subsection*{Enhanced connected correlations at large transverse field }

An important physical consequence of the constrained spreading of correlations in the large transverse-field regime is the enhancement of $C_c(r,t)$ for $t > 1/J$. To quantify this behavior, we consider the time-averaged correlations $\tilde{C}_c(r) = (1/\delta t) \int_{t_{\min}}^{t_{\max}} C_c(r,t) \, dt$, see Fig.~\ref{fig:Fig3}\textbf{E}. An analogous time average is also considered for the unconnected correlations, $\tilde{C}(r)$, built from $C(i,j,t)= \langle \hat{\sigma}_i^z \hat{\sigma}_j^z \rangle(t)$. 

For $h_x/J = 6.0$, we observe that $\tilde{C}_c(r)$ and $\tilde{C}(r)$ extend over the largest distances. Physically, this result is consistent with prethermalization to a thermal state of $\hat{H}_{ZY}$, whose temperature is defined by the energy of the initial state $\ket{\downarrow \dots \downarrow}$. In particular, this initial state corresponds to the ground state of the mean-field of $\hat{H}_{ZY}$ and is therefore expected to correspond to a low-temperature state of the short-range ZY model, which is characterized by quasi-long-range correlations \cite{harada_kosterlitz-thouless_1998,frerot_entanglement_2017}.

In contrast, for lower values of $h_x/J$, we observe a weaker build-up of $\tilde{C}_c(r)$ for $tJ > 1$. In particular, for $h_x/J = 3.5$, we observe that both the connected correlation $\tilde{C}_c(r)$ and the unconnected correlation $\tilde{C}(r)$ decay to zero at longer distances, albeit with a larger correlation length than expected from the corresponding thermal results of the QPU Hamiltonian (Methods). For $h_x/J = 2.5$, $\tilde{C}_c(r)$ decays rapidly toward zero, whereas the unconnected correlation $\tilde{C}(r)$ remains long-ranged, consistent with thermalization of the initial fully polarized state within a symmetry-broken sector.

\section*{Benchmarking tensor network approaches against QPU dynamics}

\begin{figure}
    \includegraphics{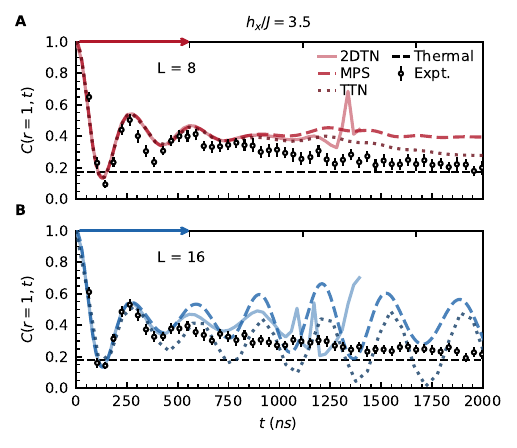}
    \caption{\textbf{Benchmarking tensor network approaches against QPU dynamics.} Post-quench dynamics of the nearest-neighbor bulk correlations for quenches to $h_x/J = 3.5$; two system sizes are considered, $L=8$ and $L=16$ respectively. In each panel, results obtained from the QPU are compared with classical simulations using 2DTN, MPS, and TTN methods, with the thermal QMC prediction shown through as a horizontal dashed line. The comparison highlights the ability of different tensor network approaches to reproduce the observed dynamics and illustrates their relative performance as the system size increases.}
    \label{fig:Fig4}
\end{figure}

Building on the previous section, where we established agreement between \textit{early-time} QPU dynamics and 2DTN simulations, we now investigate the \textit{long-time} regime where classical simulation methods become increasingly unreliable.  Here, for concreteness we focus on the intermediate regime $h_{x}/J=3.5$, i.e. the slow-relaxation regime in which classical methods are expected to struggle. Other regimes such as the DPT are also expected to be challenging~\cite{vovrosh_simulating_2026}.

Figure~\ref{fig:Fig4} compares the post-quench dynamics obtained from the QPU with simulations based on 2DTN, matrix product states (MPS), and tree tensor networks (TTN) for system sizes $L=8$ and $L=16$. At early times, all three tensor-network approaches produce quantitatively consistent results. At later times, the tensor-network predictions begin to diverge, with clear separation emerging for $tJ\gtrsim 1.5$ for $L=8$, and $tJ\gtrsim 1$ for $L=16$, reflecting the growing difficulty of accurately representing the entanglement generated during the dynamics. The disagreement between classical methods becomes comparable to, or larger than, their deviation from the QPU results, so the classical simulations no longer provide a controlled description of the long-time dynamics.

The finite bond dimensions used in tensor-network simulations place a direct restriction on the expressibility of the underlying variational ansätze, which becomes increasingly severe as the entanglement generated by the two-dimensional dynamics grows beyond what can be faithfully represented with finite computational resources. This manifests in distinct signatures in the computed dynamics. In 2DTN, this appears as late-time discontinuities, indicating a loss of numerical control. In contrast, the failure of MPS and TTN is more subtle, taking the form of pronounced oscillations absent in the QPU data. These behaviors can be understood from the structure of the ansätze: MPS is intrinsically one-dimensional, while TTN imposes a hierarchical entanglement structure that does not fully capture two-dimensional connectivity. When driven beyond their controlled regime, these methods become constrained by their built-in geometries, leading to long-time dynamics that reflect the structure of the ansätze rather than the underlying two-dimensional system.

In contrast, the QPU results remain consistent across increasing system sizes and continue to exhibit a slow decay of correlations towards the thermal values predicted by QMC. Remarkably, the intermediate regime is already qualitatively resolved in small systems accessible to state-vector simulation (see Methods), reinforcing the validity of the QPU observations at late times. While the precise temporal decay of the correlations is affected by experimental imperfections, the observed phenomenology remains robust.
Importantly, theoretical analyses of analog quantum simulators have shown that the effect of local experimental errors does not generally grow with system size, in contrast to the error growth encountered in many classical simulations, such as MPS simulations of 2D systems~\cite{trivedi_quantum_2024}. As a result, although noise modifies the quantitative value of the measured relaxation rate, it is not expected to qualitatively alter the observed dynamical behavior as the system size increases. Moreover, accurately incorporating these experimental imperfections into classical simulations currently requires averaging over many stochastic noise trajectories, each with the same computational complexity as the corresponding noiseless simulation, making the overall cost at least as demanding and often substantially greater~\cite{dalyac_noise_2026}. Consequently, reducing experimental imperfections could enable more detailed quantum dynamics studies and complement classical simulations.

\section*{Discussion and outlook}
We presented an investigation of post-quench dynamics across the phase diagram of the 2D TFIM with a 256-qubit neutral-atom QPU. We revealed a rich landscape of relaxation behaviors: in addition to regimes exhibiting rapid relaxation toward thermal expectation values (occurring on timescales of $tJ \approx 1$), we identified unexpected regimes characterized by a pronounced slowdown of thermalization and by the emergence of a prethermal plateau. An important direction for future work is to assess how these regimes depend on the choice of initial state, e.g., by comparing different initial polarizations, states with finite initial entanglement, breaking translational invariance or with an effective initial temperature. 

More generally, our results contribute to a universal problem,  the relaxation of a chaotic (nonintegrable) quantum system, which has been the focus of intense studies over recent decades. Owing to the generality of the 2D TFIM, our results can be extended across several directions, both experimentally and theoretically. At large transverse field, the emergence of an effective Floquet Hamiltonian connects naturally to previous 1D experiments with trapped ions~\cite{jurcevic_quasiparticle_2014}, and to direct simulations of the 2D XY model with dipolar Rydberg states~\cite{chen_spectroscopy_2025}, offering an attractive alternative to  generate effective $1/r^6$ XY interactions. 

At intermediate transverse fields, we  have discovered that relaxation is markedly slower than expected from the diffusive spreading of correlations associated with the onset of hydrodynamic behavior \cite{lux_hydrodynamic_2014}. This suggests that additional mechanisms may be at play, possibly linked to the breakdown of the leading-order Floquet–Magnus expansion and to critical effects associated with the nearby DPT and QPT. A similar slow-thermalizing regime has recently been investigated in the nonlocal 1D TFIM~\cite{mitra_macroproperties_2025}. An important difference is that the thermal DPT originates from the nonlocal character of the interactions, bringing the model closer to a mean-field (semiclassical) limit. Notably, this limit has also been invoked in earlier works~\cite{marino_dynamical_2022, lerose_chaotic_2018, lerose_impact_2019} to predict ferromagnetic trajectories and persistent paramagnetic oscillations, separated by an intermediate chaotic region. While the 2D TFIM studied here is far from such a semiclassical limit (due to spin-1/2 sites with quasi-local interactions), it has been conjectured~\cite{michailidis_slow_2020} that semiclassical phase structure may nevertheless leave an imprint on the quantum dynamics. It is also worth mentioning that, within an auxiliary-spin approximation, our results connect to nonequilibrium predictions in the 2D Fermi–Hubbard model~\cite{schiro_quantum_2011,michel_hubbard_2024,julia-farre_hybrid_2025}, with the transverse field playing the role of the Hubbard interaction.

Finally, such nonequilibrium dynamics is the perfect testbed to benchmark state-of-the-art numerical methods and QPU data. Our results also expose a known limitation of current state-of-the-art classical approaches based on tensor networks; in generic many-body systems, the rapid growth of entanglement makes access to late-time dynamics exponentially costly with system size. This, in turn, underscores the essential role of analog quantum simulators for probing nonequilibrium dynamics in two-dimensional, nonintegrable settings, and establishes them as a key platform to address the open questions identified here. In this context, the favorable scaling of error levels with system size~\cite{trivedi_noise_2025} is a major asset, while ongoing improvements in coherence times offer an exciting path toward systematic finite-size and finite-time scaling analyses.

\

\section*{Acknowledgments}
We thank Mourad Beji for carefully reading the manuscript and providing insightful suggestions. Pasqal Team acknowledges funding from the European Union through the project PASQuanS2.1 (HORIZON-CL4-2022-QUANTUM02-SGA, Grant Agreement 101113690). Pasqal team acknowledges AWS for their support and resources for running large-scale numerical simulations.  JT is grateful for ongoing support by the Flatiron Institute.

{\bf Note:} During the completion of this work, we became aware of related and complementary work by \emph{F. Bensch et al.} on the observation of far-from-equilibrium scaling in Rydberg (Ising) systems \cite{bensch_far_2026}. While that work focuses on the emergence of geometry-independent scaling across multiple two-dimensional lattice geometries and its description through a hierarchy of effective theoretical approaches, our work focuses on a different question: probing non-equilibrium relaxation in different regimes using a square-lattice Rydberg array.

\clearpage

\section*{Methods}
\textbf{Simulating the quantum Ising model with a Rydberg-based QPU.} The QPU is characterized by the full Rydberg Hamiltonian~\cite{browaeys_many-body_2020, scholl_quantum_2021-1}
\begin{equation}\label{eq:ham_qpu_1}
\begin{split}
    H_{\text{QPU}} = &\sum_{i<j}\frac{C_6(n)}{r_{ij}^6} \hat n_i \hat n_j+
    \frac{\hbar \Omega(t)}{2}\sum_i \hat \sigma^x_i  - \hbar \delta(t) \sum_{i}  \hat n_i.
\end{split}
\end{equation} 
Each local qubit is encoded in the $^{87}\mathrm{Rb}$ ground-state manifold $\ket{\downarrow} = \ket{5S_{1/2},\,F=2,\,m_F=2}$ and in a highly excited Rydberg state $\ket{\uparrow} = \ket{nS_{1/2},\,m_J=1/2}$, with occupation operator $\hat n_i \equiv (1 + \hat\sigma_i^z)/2$. Trapped atoms can be arranged in two-dimensional arrays with interatomic spacing $r_{ij}\sim$ a few $\mu$m and interactions between Rydberg excitations are 
governed by the van der Waals coefficient $C_6(n)$; in this work we use $n=75$. 
The ground–Rydberg transition is driven by a laser field characterized by a  time-dependent Rabi frequency $\Omega(t)$ (typically $\Omega/2\pi \approx 2~\mathrm{MHz}$) and detuning $\delta(t)$ from resonance. Under these controls, the system evolves according to the Hamiltonian $\hat H_{\rm QPU}(t)$, with pulse schedules $\Omega(t)$ and $\delta(t)$. After the evolution, the final state is measured via projective readout of the ground-state population.

In the square lattice, and using only Pauli matrices, we can rewrite it as
\begin{equation}\label{eq:ham_rydb_ising}
\begin{split}
    \hat{H}_{\text{QPU}} & = \sum_{i<j} J_{ij} \hat{\sigma}^z_i\hat{\sigma}^z_j+
    h_x(t) \sum_i \hat{\sigma}^x_i    \\
    &+\sum_{i} h_z^i (t) \hat{\sigma}^z_i+\sum_{i<j}J_{ij}.
 \end{split}
\end{equation}
Here we identified $J=C_6/(4R^6)$ as the nearest-neighbor interaction, and $J_{ij}=C_6/(4r_{ij}^6)$. We also identify $h_x(t) = \frac{\hbar\Omega(t)}{2}$, and $h^i_z(t)=-\frac{\hbar\delta(t)}{2}+\sum_{j}J_{ij}$.
Since here we are interested in the TFIM at zero longitudinal field, we choose $\delta(t)$ such that $h^i_z(t)=0$ for sites $i$ located at the center of the square array. In this way, the local edge field $h_z^i(t)$ is given by 
\begin{equation}
    h_z^i(t)=\sum_j (J_{ij} -  J_{i_\text{center}j}).
\end{equation}
The difference between the QPU and the antiferromagnetic TFIM Hamiltonian (i.e.,  $-\hat{H}_{TFIM}$, where $\hat{H}_{TFIM}$ is defined in  Eq.~\eqref{eq:ham_qpu_} of the main text) is therefore a physically irrelevant constant and local edge fields:
\begin{equation}\label{eq:edge_fields}
    \hat{H}_\textrm{QPU}-(-\hat{H}_\textrm{TFIM})=\sum_{i<j}J_{ij}+\sum_i h_z^i\hat{\sigma}^z_i.
\end{equation}
Lastly, although the Rydberg Hamiltonian maps to the antiferromagnetic Ising model, $J_{ij}>0$, the dynamics from the initial state considered in this work, i.e. the product state fully polarized in the negative $Z$-direction, maps to those of the ferromagnetic TFIM with Hamiltonian $-\hat H_{\text{QPU}}$ by time-reversal symmetry~\cite{frerot_multi-speed_2018}.
\

\textbf{Floquet prethermalization for large Rabi frequency.} Starting from the TFIM obtained from the Rydberg Hamiltonian in the previous section, we consider
\begin{equation}
\hat{H} = \sum_{i<j} J_{ij} \hat{\sigma}_i^z \hat{\sigma}_j^z + h_x \sum_i \hat{\sigma}_i^x,
\label{eq}
\end{equation}
where $J_{ij}$ reflects the underlying van-der-Waals interaction profile. We focus on the regime $h_x \gg J_{ij}$, in which the transverse field dominates the dynamics and generates rapid precession of each spin about the $x$-axis. The Ising interaction is then treated as a weak perturbation, and we derive an effective description in the rotating frame defined by the field term, followed by a time-averaging over the fast precession.

To proceed, we separate the Hamiltonian as $H = H_0 + V$, where
\begin{equation}
\hat{H}_0 = h_x \sum_i \hat{\sigma}_i^x,
\qquad
\hat{V} = \sum_{i<j} J_{ij} \hat{\sigma}_i^z \hat{\sigma}_j^z.
\end{equation}
We move to the interaction picture with respect to $H_0$, in which operators evolve as $O(t) = e^{iH_0 t} O e^{-iH_0 t}$. Since $H_0$ generates rotations about the $x$-axis, the single-spin Pauli operators transform as
\begin{equation}
\hat{\sigma}_i^x(t) \rightarrow \hat{\sigma}_i^x,
\qquad
\hat{\sigma}_i^z(t) \rightarrow \hat{\sigma}_i^z \cos(2 h_x t) + \hat{\sigma}_i^y \sin(2 h_x t),
\end{equation}
with an analogous expression for $\hat{\sigma}_i^y(t)$.

Substituting these into $V$ yields a time-dependent interaction $V(t)$, where each term $\hat{\sigma}_i^z \hat{\sigma}_j^z$ is rotated into a combination of $\hat{\sigma}^y \hat{\sigma}^y$, $\hat{\sigma}^z \hat{\sigma}^z$, and mixed $\hat{\sigma}^y \hat{\sigma}^z$ and $\hat{\sigma}^z \hat{\sigma}^y$ contributions oscillating at frequency $2 h_x$.

In the regime $h_x \gg J_{ij}$, the dynamics generated by $H_0$ is much faster than that induced by $V$, and the leading-order effective description is obtained via a secular (time-averaged) approximation. The effective Hamiltonian is given by
\begin{equation}
\hat{H}_{\mathrm{eff}} = \frac{1}{T}\int_0^T \hat{V}(t)dt,
\end{equation}
where $T = \pi/h_x$ is the period of the fast precession under $H_0$. Using $\overline{\cos^2(2h_x t)} = \overline{\sin^2(2h_x t)} = 1/2$ and $\overline{\sin(2h_x t)\cos(2h_x t)} = 0$, all rapidly oscillating mixed terms average to zero, while the diagonal contributions are retained. This yields $\hat{H}_{ZY}$ (i.e., Eq. \eqref{eq:ZYmodel})
\begin{equation}
\hat{H}_{ZY} = \sum_{i<j} \frac{J_{ij}}{2}\left(\hat{\sigma}_i^y \hat{\sigma}_j^y + \hat{\sigma}_i^z \hat{\sigma}_j^z\right).
\end{equation}

Finally, we clarify the relation between the laboratory frame and the rotating frame. The unitary transformation to the interaction picture corresponds to a time-dependent rotation about the $x$-axis, such that observables acquire explicit oscillatory dynamics at frequency $2h_x$. At stroboscopic times $t_n = n\pi/h_x$, the rotation operator reduces to a global phase, and therefore the laboratory and rotating frames coincide. 

This has two important consequences. First, the initial state considered throughout this work, $\ket{\psi_0} = \ket{\downarrow\downarrow\cdots\downarrow}$, is identical in both frames at these stroboscopic times. Second, measurements performed in the laboratory frame at times $t_n$ can be directly interpreted as measurements in the rotating frame. More generally, when measurements are performed with finite time resolution or averaged over many precession cycles, the fast oscillatory mixing between $y$ and $z$ components is washed out, and experimental observables can be directly interpreted using the effective Hamiltonian $H_{ZY}$.

\begin{figure}
    \centering
    \includegraphics[width=\linewidth]{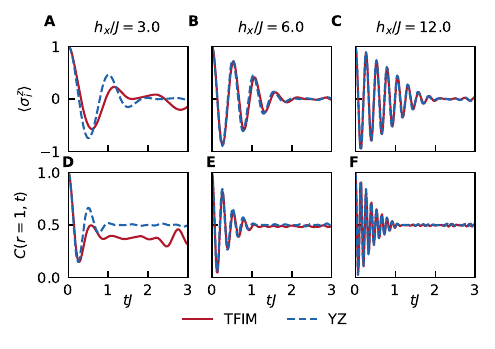}
    \caption{\textbf{Comparison between transverse-field Ising and effective ZY dynamics.}
    Time evolution following a quench from the fully polarized state $\ket{\downarrow\downarrow\cdots\downarrow}$ under the TFIM on a $4\times4$ lattice with periodic boundary conditions. Panels \textbf{A-C} show the local magnetization $\langle \hat\sigma_i^z(t)\rangle$ of a bulk site, while panels \textbf{D-F} show the nearest-neighbor correlation function $C(r=1,t)$, for transverse fields $h_x/J=3$, $6$, and $12$, respectively.}
    \label{fig:xy}
\end{figure}

To benchmark the emergence of the effective ZY-type description, we consider quenches from the fully polarized initial state in the $-z$ direction, evolving under the full TFIM on a $4\times 4$ lattice with periodic boundary conditions. In Fig.~\ref{fig:xy}, we show a comparison between the exact dynamics and the effective ZY model. The top row (panels \textbf{A-C}) displays the local magnetization $\langle \hat{\sigma}_i^z(t)\rangle$ as a function of time, $tJ$, for increasing field strengths $h_x/J = 3, 6, 12$, respectively, while the bottom row (panels \textbf{D-F}) shows the corresponding nearest-neighbor correlations. In each case, the prediction of the effective ZY model is overlaid, where we additionally include the contribution from $H_0$ in order to correctly capture the fast oscillatory dynamics in the laboratory frame. As $h_x/J$ is increased, the agreement between the full dynamics and the effective description improves systematically, with the large-field regime exhibiting excellent quantitative correspondence in both local magnetization and correlation functions, thereby confirming the validity of the rotating-frame approximation.

\

\textbf{Thermal value of $E_{cl}$ at the prethermal regime.}
In the prethermal regime, the local properties of the system are described by the thermal properties of $\hat{H}_{ZY}$ at an effective temperature set by the energy of the initial state.
In particular, this effective temperature can be obtained by matching the energy of the initial product state
$\ket{\psi(0)}=\ket{\downarrow\cdots\downarrow}$ to that of the corresponding thermal ensemble,
\begin{equation}
\bra{\psi(0)} \hat{H}_{ZY}\ket{\psi(0)}
=
\frac{{\rm Tr}\left( \hat{H}_{ZY}e^{- \hat{H}_{ZY}/(k_\text{B}T_{\mathrm{eff}})}\right)}
{\rm{Tr} \left(e^{- \hat{H}_{ZY}/(k_\text{B}T_{\mathrm{eff}})}\right)},
\end{equation}
where $k_\text{B}$ is the Boltzmann constant and $T_{\mathrm{eff}}$ is the effective temperature.

Although $T_{\mathrm{eff}}$ and the corresponding expectation values can be obtained using numerical methods, here we exploit the $U(1)$ symmetry of $\hat{H}_{\mathrm{ZY}}$ to derive an analytical expression for the thermal classical Ising energy
$E_{\mathrm{cl}}(i)_{T_{\mathrm{eff}}} = \sum_j J_{ij} \langle \hat{\sigma}_i^z \hat{\sigma}_j^z \rangle_{T_{\mathrm{eff}}}$
at $T_{\mathrm{eff}}$; where $\langle \hat O \rangle_{T_{\mathrm{eff}}}$ represents  the thermal expectation value of an operator $\hat O $ at $T_{\mathrm{eff}}$.

For the initial state considered in this work, the left-hand side is simply the classical energy of the fully polarized state and therefore evaluates to $\sum_{i<j}J_{ij}$. Furthermore, the right hand side can be simplified using the emergent $U(1)$ symmetry, which imposes rotational invariance in the $yz$ plane, $\langle\hat{\sigma}_i^z\hat{\sigma}_j^z\rangle_{T_{\mathrm{eff}}} \equiv \langle\hat{\sigma}_i^y\hat{\sigma}_j^y\rangle_{T_{\mathrm{eff}}}$. Consequently, the thermal expectation value of the effective Hamiltonian
reduces to $\langle H_{ZY}\rangle_{T_{\mathrm{eff}}} = 2 \sum_j J_{ij} \langle \hat{\sigma}_i^z \hat{\sigma}_j^z \rangle_{T_{\mathrm{eff}}}$, which is precisely the thermal value of $E_{\mathrm{cl}}(i)$; hence,
\begin{equation}
    E_{\mathrm{cl}}(i)_{T_{\mathrm{eff}}} = \frac{1}{2}\sum_{i<j}J_{ij}.
\end{equation}
This provides a direct benchmark for the existence of the prethermal plateau, allowing us to verify its presence across all system sizes without requiring a full dynamical simulation.

\

\textbf{Exact numerical simulations of the relaxation dynamics}.  We perform exact numerical simulations of the unitary relaxation dynamics using state vector simulations on $L=5$ square lattices. Remarkably, the three dynamical regimes identified in the main text are already clearly visible at small system sizes, and the exact approach allows us to more sharply characterize their boundaries in parameter space and time.

\begin{figure}
    \centering
    \includegraphics[width=1\linewidth]{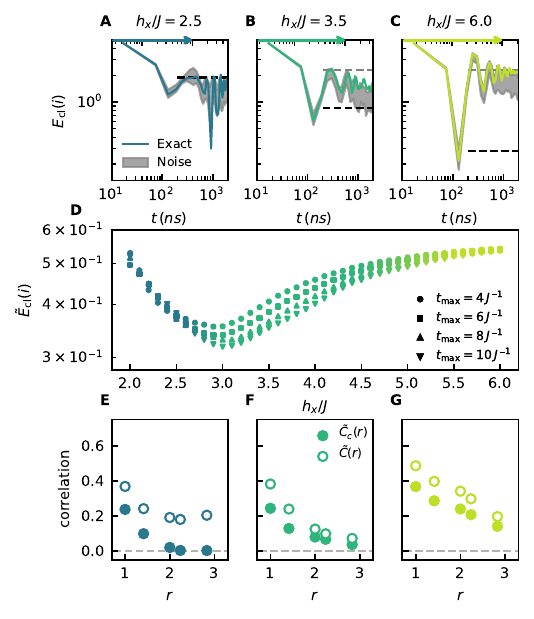}
     \caption{\textbf{Exact numerical simulations of relaxation dynamics in finite systems.} \textbf{A–C} Time evolution of $E_{\text{cl}}(i)$ for a bulk site in a $5\times5$ lattice illustrating three distinct dynamical regimes are already present at small system sizes: fast thermalization (\textbf{A}), slow relaxation (\textbf{B}), and prethermalization (\textbf{C}). In addition, we include results for a physically motivated noise model, obtained using MPS simulations. \textbf{D} Time-integrated results of the classical bulk energy, $\tilde{E}_{\text{cl}}(i)$, with $t_{\text{min}} = 0$ and evaluated for varying $t_{\text{max}}$. \textbf{E–G} Time-integrated results of both long-range correlations and connected correlation functions compared across the three regimes between $t_{\text{min}} = 1$ and $t_{\text{max}}=2$.}
    \label{fig:Fig_noise}
\end{figure}

To illustrate this, Fig.~\ref{fig:Fig_noise} \textbf{A}–\textbf{C} shows representative dynamics of $E_{\text{cl}}(i,t)$ for a bulk site for the three regimes: fast thermalization, slow relaxation, and prethermalization. In the fast thermalization regime, observables rapidly approach their thermal values predicted by QMC, up to finite-size induced oscillations around the steady state. In contrast, the slow relaxation regime exhibits markedly delayed convergence, with persistent transient dynamics over extended timescales. Finally, in the prethermalization regime, the system rapidly approaches a long-lived non-thermal plateau, consistent with prethermal behavior, and remains steady for the full simulation timescales.

To further quantify the crossover between these regimes, we consider the time-integrated observable $\tilde{E}_{\mathrm{cl}}(i)=(1/\delta t)\int_{t_{\min}}^{t_{\max}} E_{\mathrm{cl}}(i,t)dt$, where $\delta t =t_{\max}-t_{\min}$, as a function of $h_x$ and varying upper integration bound $t_{\max}$. As shown in Fig.~\ref{fig:Fig_noise} \textbf{D}, in both the fast thermalization and prethermalization regimes the integrated quantity quickly saturates and becomes effectively independent of $t_{\max}$, reflecting rapid equilibration to either thermal or prethermal steady values. In contrast, within the slow relaxation regime, the integral exhibits a pronounced dependence on $t_{\max}$, systematically decreasing towards the thermal expectation as longer time windows are included. This behavior provides a quantitative diagnostic distinguishing slow relaxation from both fast thermalization and prethermal trapping.

Finally, in Fig.~\ref{fig:Fig_noise} \textbf{E}-\textbf{G} we compare time-integrated long-range connected correlation functions across the three regimes. We find that the qualitative structure of correlations is already well captured at small system sizes, including the emergence of extended spatial correlations in the prethermalization regime. In particular, the prethermal regime is characterized by quasi long-range connected correlations, consistent with the presence of quasi-stationary, pre-thermal states.

\

\textbf{Details on the noise model for all three regimes.} The ability of the QPU to simulate the unitary dynamics governed by the Hamiltonian \eqref{eq:ham_qpu_1} is ultimately limited by the different noise sources present in the device. These can be broadly separated into \textit{(i)} state preparation and measurement errors (SPAM), \textit{(ii)} static offsets and fluctuations in the Hamiltonian parameters, and \textit{(iii)} effective dephasing channels associated to high-frequency laser phase noise and to scattering from the intermediate state used to couple $\ket{\downarrow}$ and $\ket{\uparrow}$ via a two-photon transition~\cite{dalyac_noise_2026}. The strength of these different sources is estimated via independent calibrations. Measurement errors represent classical noise in the QPU bitstrings, which is taken into account when using them to compute single or two-body observables. The remaining errors can be fed into a noisy emulator of the many-body dynamics to assess their impact in the physics investigated in this work. 

In Fig.~\ref{fig:Fig_noise}, we report on this analysis for $L=5$ square lattices in the three regimes considered in the main text. At this small system size, our MPS emulator would cover the full Hilbert space already at $\chi=4096$, and we certified that $\chi=512$ faithfully reproduces exact state-vector emulations for the dynamics under consideration. For each value $h_x/J$, we compare ideal emulation of the Hamiltonian unitary dynamics, the simulation using the QPU, and the noisy emulation. On the one hand, the compatibility between the QPU data and the noisy emulator is in agreement with our noise modeling, based on local (uncorrelated) noise sources. According to this picture, errors in local observables should mainly scale with time but not with system size~\cite{trivedi_noise_2025}. On the other hand, we observe that, as generally expected in many-body dynamics under noise, correlations exhibit a decay compared with the ideal emulator, and on a time scale ($t\gtrsim 1~\mu\textrm{s}$) shorter than noninteracting single-atom coherence ($T_2\approx 10\mu s$ in the device). 

Although the huge difference in system sizes prevents a direct comparison of this noise analysis ($L=5$) with the main text QPU data ($L=16$), the former provides us with qualitative insights of noise effects in different Hamiltonian regimes for the timescale $t\leq 2\,\mu\textrm{s}$ considered in the main text. At the largest $h_x/J=6$, the noise model predicts a drift below the prethermal plateau. Interestingly, this effect seems to be also present at $L=16$ (Fig.~\ref{fig:Fig2}\textbf{C}). At the intermediate $h_x/J=3.5$, we observe that the many-body decay of correlations is further enhanced by noise effects. Thus, while in the current work we showed that the QPU is sensitive to the different regimes at $L=16$, incrementing the coherence time of the device is a clear prospect for hardware improvement, to disentangle the role of many-body effects and noise. Finally, for $h_x/J=2.5$ the noise analysis predicts a shift towards smaller correlations, compared to the ideal emulation. Although the $L=16$ data in this regime (Fig.~\ref{fig:Fig2}\textbf{C}) thermalize to the predicted QMC value, we note that here the larger interaction value enhances edge effects, making the comparison between $L=5$ and $L=16$ results even more challenging.

\

\begin{figure}
    \centering
    \includegraphics[width=1.0\linewidth]{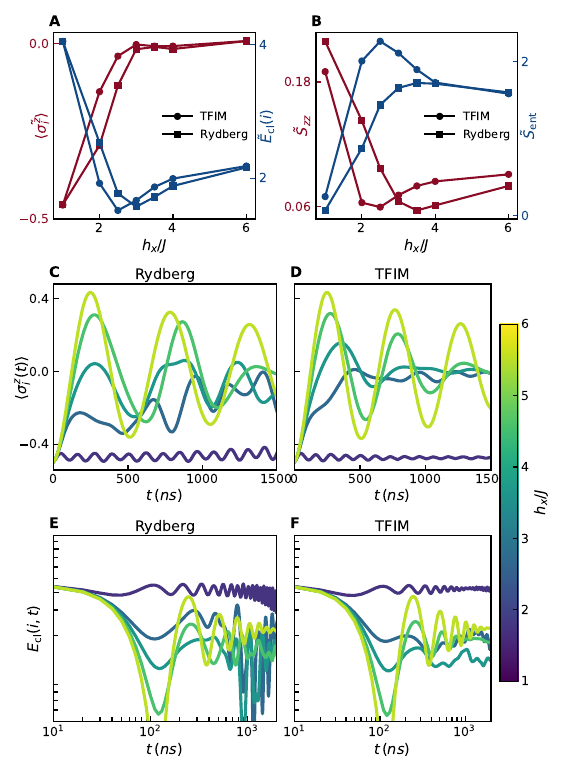}
    \caption{\textbf{Comparison between dynamics generated by the Rydberg Hamiltonian (QPU) and the TFIM on a $5 \times 5$ square lattice.}
\textbf{A-B} Time-averaged observables over the interval $tJ \in [0,2]$. Panel \textbf{A} shows the central-site local magnetization and classical energy, while panel \textbf{B} shows the global structure factor and entanglement entropy.
\textbf{C-D} Time evolution of the central-site local magnetization for different values of $h_x/J$, comparing the Rydberg Hamiltonian in \textbf{C} with the TFIM in \textbf{D}.
\textbf{E-F} Time evolution of the central-site classical energy for the same values of $h_x/J$, comparing the Rydberg Hamiltonian in \textbf{E} with the TFIM in \textbf{F}.
The colorbar indicates the color coding for $h_x/J = 1, 2, 2.5, 3.5,$ and $6$ used in panels \textbf{C-F}.}
    \label{fig:fig_ising}
\end{figure}

\textbf{Benchmark of edge effects in the QPU Rydberg Hamiltonian dynamics.}
We also use exact numerical simulations on a small $L=5$ lattice to compare the QPU Rydberg Hamiltonian with the ideal TFIM of Eq.~\eqref{eq:ham_qpu_}, and thereby benchmark the effects of the residual edge fields 
of Eq.~\eqref{eq:edge_fields}. Figure~\ref{fig:fig_ising}\textbf{A} shows the comparison of time-averaged local quantities, namely the local magnetization and the classical energy. The two
Hamiltonians display similar qualitative behavior, but the Rydberg
Hamiltonian exhibits a clear shift of the crossover features toward larger
values of $h_x/J$. This shift is even more apparent in the time-averaged global quantities shown in
Fig.~\ref{fig:fig_ising}\textbf{B}. In particular, we show the structure factor
$S_{zz}(t)=\frac{1}{L^4}\sum_{ij}\langle \hat{\sigma}_i^z\hat{\sigma}_j^z\rangle$ and the half-cut entanglement entropy $S_{\rm ent}(t)$, the latter computed using quasi-exact MPS
simulations with bond dimension $\chi=1024$. These results show that the explicit
breaking of the $\mathbb{Z}_2$ Ising symmetry by the residual edge fields
broadens and shifts the finite-size and finite-time signatures of the
dynamical phase transition.
The latter can also be inferred from the 
time-resolved dynamics of the local magnetization, as shown in Figs.~\ref{fig:fig_ising}\textbf{C-D}: while in the TFIM magnetization trajectories (Fig.~\ref{fig:fig_ising}\textbf{D}) one can identify a special curve separating FM and PM trajectories around $h_x/J\approx 2.5$, reminiscent of a separatrix in parent mean-field models, the Rydberg Hamiltonian (Fig.~\ref{fig:fig_ising}\textbf{C}) exhibits a less clean behavior in the same transition regime. Similarly, the local classical energy curves, shown in Figs.~\ref{fig:fig_ising}\textbf{E-F}, are also consistent with extra edge-induced oscillations in the Rydberg model, which nevertheless exhibits the same qualitative relaxation regimes as the TFIM. As a final remark, we note that, while these edge-induced discrepancies between the Rydberg and TFIM Hamiltonians are expected to be mitigated in the $L=16$ QPU simulations presented in the main text, due to a larger bulk-to-edge ratio, their impact on critical modes with large correlation lengths might remain relevant, as well as their interplay with the noise sources discussed in the previous section. Mitigating edge effects using local detuning maps is therefore a promising route to sharpen the characterization of the dynamical transition in finite-size arrays. In particular, to better distinguish its associated critical slowing down from the emergence of prethermal Floquet dynamics at larger but nearby $h_x/J$.

\

\textbf{Quantum Monte Carlo simulations.}
We use quantum Monte Carlo (QMC)  based on stochastic series expansion (SSE) to simulate thermal properties of the Hamiltonian $ \hat H_{QPU}$ (as defined in Eq.~\eqref{eq:ham_rydb_ising}).  This algorithm relies on a classical representation obtained from a Taylor expansion of the partition function and is implemented using both local and cluster updates \cite{sandvik_stochastic_2003,merali_stochastic_2024}. 
Consistent with the QPU setup, we perform simulations  in a square atomic array, containing $N = L \times L$ qubits. In all simulations, we use $N_{\mathrm{therm}} = 5 \times 10^4$ Monte Carlo steps for thermalization, and observables are evaluated using $N_{\mathrm{meas}} = 10^7$ successive measurements. A binning analysis is then used to estimate errors. 

\

\begin{figure}
    \centering
    \includegraphics[width=1.0\linewidth]{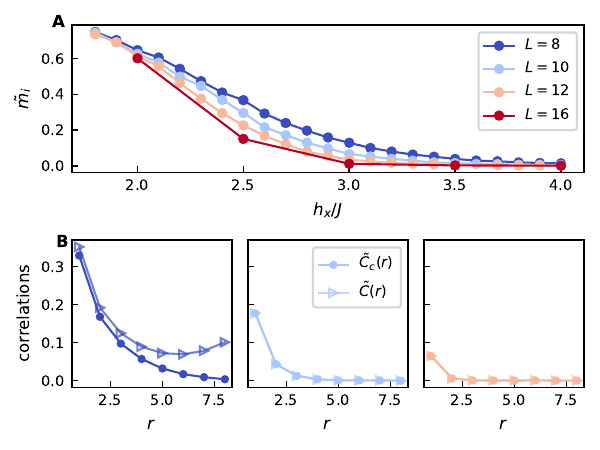}
    \caption{\textbf{Thermal equilibrium results at effective temperatures fixed by the quenches.}  We show thermal  observables computed at different effective temperatures $T_{\mathrm{eff}}(h_x/J)$. In particular, we consider (A) the $h_x/J$-dependence of the bulk magnetization, $\langle \hat{\sigma}^{z}_{i}\rangle$,
    and (B) the connected and unconnected correlations obtained for the three representative values of $h_x/J$  considered in the main text: $h_x/J = 2.5$ (left panel), $h_x/J = 3.5$ (central panel) and $h_x/J = 6.0$ (right panel). }
    \label{fig:Fig_qmc}
\end{figure}

\textbf{ Thermal equilibrium properties of $\hat{H}_{\mathrm{QPU}}$ and further comparison with QPU results.}
We use the QMC-SSE approach to compute the effective temperature defined by the energy injection in the different quenches considered in the main text. 
 In particular, we assume energy conservation and match the initial energy (i.e., the one associated with the state $ \ket{\psi(0)} = \ket{\downarrow \dots \downarrow}$) with that of a
thermal ensemble at an effective temperature $T_{\mathrm{eff}}$,
\begin{equation}
\bra{\psi(0)} \hat{H}_{\mathrm{QPU}} \ket{\psi(0)} =
\frac{Tr\left( \hat{H}_{\mathrm{QPU}} e^{- \hat{H}_{\mathrm{QPU}}/(k_\text{B}T_{\mathrm{eff}})} \right)}{Tr\left( e^{- \hat{H}_{\mathrm{QPU}}/(k_\text{B}T_{\mathrm{eff}})} \right)},
\label{eq:eth_sm}
\end{equation}
where $\hat{H}_{\mathrm{QPU}}$ is defined in Eq.~\eqref{eq:ham_rydb_ising}.

Solving Eq.~\eqref{eq:eth_sm} for different values of $h_x$ allows us to obtain $T_{\mathrm{eff}}(h_x/J)$. We note that although the initial energy is independent of $h_x$, i.e., $E(0) = \bra{\psi(0)} \hat{H}_{\mathrm{QPU}} \ket{\psi(0)} = 0$, the thermal energy depends on this parameter. In practice, to obtain $T_{\mathrm{eff}}(h_x/J)$, we use QMC simulations to determine the $T_{\mathrm{eff}}$ at which the thermal energy matches $E(0)$ for different values of $h_x$. Examples of $T_{\mathrm{eff}}/J$ are shown in Fig.~\ref{fig:Fig1}C for $L=16$. 
Interestingly, the solution of Eq.~\eqref{eq:eth_sm} for the initial state has a negative temperature, $T_{\mathrm{eff}} < 0$. This can also be physically interpreted as a Hamiltonian with ferromagnetic interactions, i.e., by considering $-\hat{H}_{\mathrm{QPU}}$. 
Another notable aspect of these results is that $T_{\mathrm{eff}}/J$ crosses the thermal transition of the ferromagnetic state at $T_{\mathrm{eff}}(h_t^*/J)$, which corresponds to the thermal estimate of the DPT. We also note that the estimate of $T_{\mathrm{eff}}(h_t^*/J)$ for $\hat{H}_{\mathrm{QPU}}$ is shifted toward larger values of $h_x/J$ compared with the estimates for the TFIM~\cite{katschke_finite-temperature_2026}.

In Fig.~\ref{fig:Fig_qmc}\textbf{A}, we consider the thermal expectation value of the bulk magnetization, i.e., the magnetization $\langle \hat{\sigma}^z_i \rangle$ averaged over the central-most sites, for different values of $T_{\mathrm{eff}}(h_x/J)$.
We emphasize that the presence of a residual longitudinal field in $\hat{H}_{\mathrm{QPU}}$ weakly breaks the $\mathrm{Z}_2$ symmetry of the conventional TFIM, due to longitudinal fields at the edges of the square array. In the thermodynamic limit, $L \to \infty$, such longitudinal fields are expected to be negligible in the majority of the system, and a thermal transition is expected to take place, as occurs for the TFIM.
In fact, we observe that $\langle \hat{\sigma}^z_i \rangle \to 0$ for $h_x/J > 2.5$ at $L = 16$, signaling the onset of the thermal phase transition. We therefore consider $h_t^*/J \approx 2.6$ as a reference for quenches in the vicinity of the DPT for the system size $L = 16$ considered in this work. This estimate is expected to shift to smaller values of $h_x/J$ for larger system sizes, in agreement with the predictions for the TFIM~\cite{katschke_finite-temperature_2026}.

For comparison with the results shown in Fig.~\ref{fig:Fig3}\textbf{E}, we also show in Fig.~\ref{fig:Fig_qmc}\textbf{B} the corresponding correlations for the three representative values of $h_x/J$ considered in the main text.

\

\textbf{Two-dimensional tensor networks.}
We represent the time-evolved many-body wavefunction $\ket{\psi}$ as a tensor network state (TNS) whose geometry matches that of the $L\times L$ qubit array, placing one tensor per qubit and connecting nearest neighbors by virtual indices of maximum size $\chi$ (the bond dimension). For a $L \times L$ square lattice the coordination number is $z=4$, so the memory footprint scales as $\mathcal{O}(L^{2}\chi^z)$, linear in the number of qubits. The time evolution under $\hat H_{\mathrm{QPU}}$ is implemented through a first-order Trotter--Suzuki decomposition into layers of non-overlapping gates with time slice $dt J= 0.01$, obtained from an edge coloring of the lattice.

Gates, not to be confused with quantum gates, are applied sequentially. One-qubit gates act exactly without altering the bond dimension. Nearest-neighbor two-qubit gates are applied to the relevant pair of tensors, and the resulting object is truncated back to dimension $\chi$ via a singular value decomposition (SVD) conditioned on belief propagation (BP) message tensors incident to the pair~\cite{jiang_accurate_2008,alkabetz_tensor_2021,tindall_gauging_2023}. The longer-range couplings arising from the $1/r^6$ interaction are handled at the next-nearest-neighbor level by applying the corresponding gate as a three-site matrix product operator (MPO) across the relevant patch; the bonds within this patch are then truncated back to $\chi$ under the BP approximation, using the BP messages incident to the three-site region as the environment. Increasing $\chi$ systematically reduces the truncation error on all two-qubit gate applications, and the TNS representation becomes exact when no singular values are discarded. To keep the environments up to date, the BP message tensors are refreshed between Trotter layers. The total cost of the evolution scales as $\mathcal{O}(t\chi^{z+1}L^{2})$, where $t$ is the total time of the simulation.

To extract observables, we first truncate the TNS down, under the BP approximation, to a more affordable bond dimension $\chi' < \chi$ \cite{tindall_dynamics_2026}. We then partition the network by rows and sweep a boundary MPS of bond dimension $R$ through the norm network $\langle\psi|\psi\rangle$ via variational MPS--MPO fitting~\cite{verstraete_renormalization_2004,lubasch_algorithms_2014, rudolph_simulating_2025}. The connected correlators $C_c(\delta_n)$ between sites along a given row are then evaluated using the MPS incident to that row, which approximates the contraction of the remainder of the network. The contraction accuracy typically improves with $R$ (it is exact as $R\to\infty$), at a cost of $\mathcal{O}(\chi^{6}R^{2}L^{2}) + \mathcal{O}(\chi^{4}R^{3}L^{2})$ when $R \geq \chi$. All tensor contractions are GPU-accelerated, which provide massive speedups for 2D TNS calculations \cite{rudolph_simulating_2025}, and the simulations are done using the \textit{TensorNetworkQuantumSimulator.jl} library \cite{tindall_tensornetworkquantumsimulatorjl_2025}. All $2$D TNS data presented in the main text of the manuscript is for $\chi = 40$, $\chi' = 24$ and $R = 48$, which we find provides a good tradeoff between efficiency and accuracy for the considered times. 

\

\textbf{Matrix-product-states.}
We use a matrix product state (MPS) framework implemented in emu-mps \cite{bidzhiev_efficient_2025}, with time evolution performed using a two-site time-dependent variational principle (2-site TDVP) algorithm. The Hamiltonian is represented in matrix product operator (MPO) form, and time evolution is carried out within the MPS manifold with adaptive bond dimension growth enabled by the two-site update structure. After each local update, tensors are re-orthogonalized and truncated via singular value decomposition (SVD) with a prescribed cutoff. The corresponding memory footprint and FLOPS scale as $\mathcal{O}(L^{3}\chi^2)$ and $\mathcal{O}(L^{3}\chi^3)$ respectively.

For two-dimensional square lattice simulations, the system is mapped onto a one-dimensional MPS ordering using a snake (serpentine) path. The mapping proceeds row by row, starting from the bottom-left corner and traversing each row from left to right. All simulations are performed using GPU acceleration. Simulations were run on NVIDIA H200 GPUs with bond dimensions $\chi=1000$ and $\chi=600$ for system sizes $L=8$ and $L=16$ respectively. A detailed analysis of resource requirements with increasing system size can be found in Ref.~\cite{vovrosh_resource_2026}.

\

\textbf{Tree-tensor networks.}
The physical state is approximated using a binary tree architecture, which allows one to naturally encode states on lattices with linear dimensions of the form $d=2^n, \ n\in\mathbb{N}$; lattices not following this rule are embedded into the next larger lattice of that form. We furthermore alternate the connection direction between two layers of the tree in order to better accommodate the structure of the interactions in two-dimensional systems. 
The time evolution is performed via the single-site TDVP using TTN.jl~\cite{tausendpfund_ttnjl_2024}. While the single-site algorithm conserves the energy per construction, it does not allow for the bond dimension to be dynamically adapted during the simulation. As such, the maximum desired bond dimension has to be set for the initial state, which is achieved by padding the tensors with $0$'s. The application of the effective Hamiltonian is performed via local sums, which for two-dimensional systems significantly reduces the scaling compared to an MPO encoding. 

To reduce the memory requirements, the tensors are stored predominantly in RAM, and only loaded into GPU memory when required by the algorithm. The main memory bottleneck lies in the Lanczos algorithm, which is used to update each tensor sequentially via the associated environment, typically requiring one to store around 20-30 Krylov vectors in order to compute all coefficients and subsequently accumulate the final result. We circumvent this problem by utilizing a two-pass adaptation of the Lanczos algorithm, in which the coefficients are calculated in the first pass, discarding the Krylov vectors, which are then recalculated in the second pass to reconstruct the final result. This reduces the number of Krylov vectors saved in GPU memory to around 5, albeit at the cost of an increase in runtime by a factor of around two. The corresponding memory footprint and FLOPS scale as $\mathcal{O}(L^{2}\log(L^2)\chi^4)$ and $\mathcal{O}(L^{2}\log(L^2)\chi^3)$ respectively.

The simulations are performed on NVIDIA A100 and H100 GPUs, both of which allow for 80GB of VRAM. The largest bond dimension is set according to the lattice size, such that the available GPU memory is fully utilized; in particular, we use $\chi=724$ and $\chi=512$ for system sizes $L=8$ and $L=16$ respectively.

\

\textbf{State preparation and measurement error mitigation.} In all results presented in this work, false positive $\varepsilon$ and false negative $\varepsilon^\prime$ detection errors ~\cite{leseleuc_analysis_2018} are corrected at the level of single-body magnetization and two-body correlation functions using independently measured error rates. In particular, we find that $\varepsilon = 1\%$ and $\varepsilon^\prime = 3\%$. The corrected magnetization $M(i,t)=\langle\hat\sigma^z_i\rangle(t)$ per site is obtained from the measured value $\bar{M}(i,t)$ as
\begin{equation}
M(i,t) = \frac{\bar{M}(i,t) - (\varepsilon^\prime - \varepsilon)}{1 - \varepsilon - \varepsilon^\prime},
\end{equation}
and similarly for corrected correlations $C(i,j,t)=\langle\hat\sigma^z_i\hat\sigma^z_j\rangle(t)$ from measured correlations $\bar{C}(i,j,t)$,
\begin{equation}
C(i,j,t) = \frac{\bar{C}(i,j,t) - (\varepsilon^\prime - \varepsilon)(\bar{M}(i,t) + \bar{M}(j,t)) - (\varepsilon^\prime - \varepsilon)^2}{(1 - \varepsilon - \varepsilon^\prime)^2}.
\end{equation}

\bibliography{quanta_experiment_paper}
\clearpage
\onecolumngrid

\end{document}